\documentstyle[sprocl,psfig]{article}
\def\Journal#1#2#3#4{{#1} {\bf #2}, #3 (#4)}

\def\bk{{\bf k}}
\def\bsig{\hbox{\boldmath$\sigma$}}
\def\lapprox{\mathrel{\mathop
  {\hbox{\lower0.5ex\hbox{$\sim$}\kern-0.8em\lower-0.7ex\hbox{$<$}}}}}
\def\gapprox{\mathrel{\mathop
  {\hbox{\lower0.5ex\hbox{$\sim$}\kern-0.8em\lower-0.7ex\hbox{$>$}}}}}

\begin{document}

\vbox to 0pt{\vskip -2.5cm\raggedright
\noindent\tt Contribution to the Proceedings of the 
XVII International Conference on Neutrino Physics and Astrophysics
(Neutrino 96), Helsinki, Finland, June 13-19, 1996.\par\vfil}

\vskip 0pt minus 1000pt

\title{SUPERNOVA NEUTRINO OPACITIES}

\author{G.G.~RAFFELT}

\address{Max-Planck-Institut f\"ur Physik (Werner-Heisenberg-Institut)\\
F\"ohringer Ring 6, 80805 M\"unchen, Germany}

\maketitle

\abstracts{In a supernova (SN) core during infall and after collapse,
neutrinos are trapped by their interactions with nucleons.  For the
axial-vector current interactions it is not enough to include static
particle-particle correlations. Even the single-nucleon spin
autocorrelation reduces the scattering cross section
dramatically. Therefore, the dynamical structure of the spin-density
structure function has an important impact on neutrino transport in
supernovae. At the present time, SN neutrino opacities cannot be
calculated to a controlled degree of precision.}


\section{Introduction}

The collapsed cores of supernovae provide the only known environments
which are so dense and hot that neutrinos are trapped. The transport
of energy and lepton number is governed by the neutrino transport
coefficients, i.e.\ by their effective scattering and absorption cross
sections with the medium constituents.  While electrons cannot be
entirely ignored, it is the interaction with nucleons which is thought
to dominate the neutrino opacities.

The scattering cross sections can be dramatically modified by
correlation effects. In the context of SN physics, the best-known
example is the coherent enhancement of the neutral-current scattering
cross section off large nuclei which is instrumental for the
lepton-number trapping in a SN core during the infall
phase~\cite{Freedman}. Recently it has been recognized that this
enhancement is partly undone by the nucleus-nucleus anticorrelations
caused by their Coulomb repulsion~\cite{Leinson,Horowitz}.

After collapse when the shock wave has dissociated the remaining layer
of large nuclei, the neutrino opacities are governed by single protons
and neutrons.  In the nonrelativistic limit the neutral-current
scattering cross section is $(G_{\rm
F}^2/4\pi)\,(C_V^2+3C_A^2)\,E_\nu^2$ where $C_V$ and $C_A$ are the
axial-vector current couplings; $C_V^2=1$ for neutrons and
$C_V^2\approx0$ for protons. Further, $C_A^2\approx 1$ for both
protons and neutrons, but the exact values in a nuclear medium are not
known~\cite{RaffeltSeckel}. What is relevant for neutrino transport is
actually an angular average weighted with $(1-\cos\theta)$, leading to
the so-called transport cross section $(G_{\rm
F}^2/3\pi)\,(C_V^2+5C_A^2)\,E_\nu^2$.  Either way, neutrino scattering
is now dominated by the axial-vector current interactions because in
this medium of single nucleons the coherent enhancement of the
vector-current cross section no longer operates.

\eject
 
Nucleons interact by a spin-dependent force so that one expects
spin-spin correlations. In nuclei, the nucleon spins tend to pair off,
leading to a coherent reduction of the axial-vector current cross
section, in contrast with the coherent enhancement of the vector
current part. For the hot nuclear matter of a nascent neutron star,
the static spin-spin correlations have been studied 
once~\cite{Sawyer89}, indicating a significant pairing effect and thus a
suppression of the effective scattering cross section. 

More recently, it has been recognized that there is also a
very significant suppression effect from the dynamical structure of
the spin correlation function. The basic idea is that the target spin
evolves ``during'' the collision with a neutrino. As a result, the
neutrino ``sees'' a reduced average spin and thus scatters less
efficiently. This reduction effect has been the subject of a series of
papers involving the present
author~\cite{RaffeltSeckel,KJR,JKRS,RSS,RaffeltStrobel}, and has also
been studied by other groups~\cite{Sawyer95,Sigl}.
In the nonrelativistic limit, there is no corresponding vector-current
effect because the zeroth component of the vector current (the charge
density) is conserved. A charge which is being kicked around cannot
``cancel itself'', in contrast with a spin when it flips back and
forth due to collisions.

\section{The Spin-Density Structure Function}

Sawyer~\cite{Sawyer95} has shown that the cross-section reduction by
nucleon spin fluctuations can be derived perturbatively by calculating
the $\nu N$ scattering cross section in the presence of a
spin-dependent external potential for the nucleons~$N$. However, a
deeper understanding is achieved in the language of linear-response
theory. 

I assume that only a single species of nucleons is involved.  The
neutral axial-vector current neutrino interaction is based on the
Hamiltonian density $(C_A G_{\rm F}/2\sqrt2)
\overline\psi_N\gamma_\mu\gamma_5\psi_N\,
\overline\psi_\nu\gamma^\mu(1-\gamma_5)\psi_\nu$.  If I further focus
on an isotropic, nonrelativistic, nondegenerate medium of baryon
density $n_B$ and temperature $T$, the axial-current transition rate
for a neutrino of four momentum $(E_1,\bk_1)$ to $(E_2,\bk_2)$ is
found to be $(C_A^2 G_{\rm
F}^2/4)\,n_B(3-\cos\theta)\,S_\sigma(\omega,\bk)$ with $\theta$ the
scattering angle and $(\omega,\bk)=(E_1-E_2,\bk_1-\bk_2)$ the 
energy-momentum transfer. Here, the dynamical spin-density structure
function is defined by
\begin{equation}
S_\sigma(\omega,\bk)=\frac{4}{3n_B}\int_{-\infty}^{+\infty}
dt\,e^{i\omega t}\langle\bsig(t,\bk)\cdot\bsig(0,-\bk)\rangle\,,
\label{eq:1}
\end{equation}
where $\bsig(t,\bk)$ is the spatial Fourier transform at time $t$ of
the nucleon spin-density operator. The expectation value
$\langle\ldots\rangle$ is taken over a thermal ensemble so that
detailed balance $S_\sigma(\omega,\bk)=S_\sigma(-\omega,-\bk)\,
e^{\omega/T}$ is satisfied. 

If one neglects the momentum transfer from neutrinos to
nonrelativistic nucleons (long-wavelength approximation), only
$S_\sigma(\omega)=\lim_{\bk\to0}S_\sigma(\omega,\bk)$ is used. After an
angular integration the axial-current scattering cross section is
\begin{equation}
\frac{d\sigma_A}{d E_2}=\frac{3 C_A^2 G_{\rm F}^2}{4\pi}
\,\frac{E_2^2\,S_\sigma(E_1-E_2)}{2\pi}\,,
\label{eq:2}
\end{equation}
where $S_\sigma(\omega)=(4/3n_B) \int_{-\infty}^{+\infty} 
dt\,e^{i\omega t} \langle \bsig(t)\cdot\bsig(0)\rangle$ with 
$\bsig(t)\equiv\bsig(t,0)$ is the total spin operator for the complete
ensemble of nucleons. 
In a noninteracting medium the nucleon spins do not evolve, and so
$\bsig(t)=\bsig(0)$. Then the time integration yields
$S_\sigma(\omega)=2\pi\delta(\omega)$. In Eq.~(\ref{eq:2}) one thus
recovers the standard expression for the scattering cross section
where recoil effects have been ignored. 

Nucleon-nucleon collisions with a spin-dependent force cause a
nontrivial evolution of $\bsig(t)$. Still, in 
$\int_{-\infty}^{+\infty}d\omega\,S_\sigma(\omega)$ the $e^{i\omega t}$
factor gives one $2\pi\delta(t)$. Therefore, one finds that the
``sum'' $\int_{-\infty}^{+\infty} d\omega\, S_\sigma(\omega)/2\pi
=(4/3n_B)\langle\bsig(0)\cdot\bsig(0)\rangle$ is independent of the
time evolution of $\bsig(t)$. In the present discussion I ignore $NN$
correlations so that one finds the simple sum rule
$\int_{-\infty}^{+\infty} d\omega\, S_\sigma(\omega)=2\pi$.

In an interacting medium the nucleon spins ``forget'' their initial
orientation after a timescale $\Gamma_\sigma^{-1}$ where
$\Gamma_\sigma$ is the spin-fluctuation rate due to collisions. It
roughly represents the width of the Fourier-transformed nucleon spin
autocorrelation function $S_\sigma(\omega)$. For
$\omega\gapprox\Gamma_\sigma$ one can
calculate $S_\sigma(\omega)$ perturbatively. It is easiest to compute
$d\sigma_A/d E_2$ directly for the $\nu N N\to N N\nu$ process
according to the usual Feynman rules, including a $NN$ interaction
potential. One can then extract $S_\sigma(\omega)$ by removing
coupling constants and phase-space factors according to
Eq.~(\ref{eq:2}). This calculation is closely related to that of the
bremsstrahlung emission of $\overline\nu\nu$ pairs in $NN$ collisions.
Generically, the result is of the form
$S^{\rm brems}_\sigma(\omega)=(\Gamma_\sigma/\omega^2)\,s(\omega/T)\,
b(\omega/T)$ where $b(x)=1$ for $x>0$ and $e^x$ for $x<0$ ensures the
detailed-balance condition. The function $s(x)$ is even, slowly varying,
and normalized according to $s(0)=1$. For large energy transfers it
represents details about the assumed $NN$ interaction potential. 

A perturbative calculation cannot reveal $S_\sigma(\omega)$ in the
$\omega\lapprox\Gamma_\sigma$ regime where multiple-scattering effects
dominate. Motivated by the classical
analogy of a spin vector kicked around by a random force one may use
the Lorentzian ansatz
\begin{equation}
S_\sigma(\omega)=\frac{\Gamma_\sigma}{\omega^2+\Gamma^2/4}\,
s(\omega/T)\,b(\omega/T)\,, 
\label{eq:3}
\end{equation}
where $\Gamma$ is chosen such that the sum-rule is fulfilled. 

\section{Cross-Section Reduction}

\begin{figure}
\center\leavevmode
\psfig{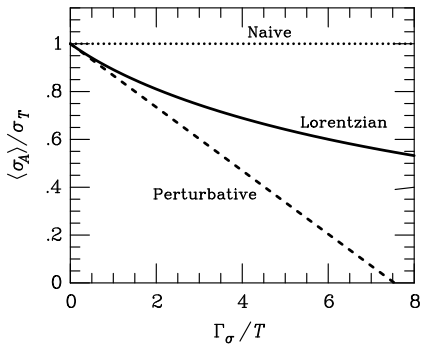}
\caption{Variation of the thermally averaged axial-current neutrino
scattering cross section on nucleons as a function of the nucleon 
spin-fluctuation rate.
\label{fig:reduction}}
\end{figure}

One may then proceed to calculate the total neutrino scattering cross
section, averaged over a thermal distribution of neutrino
energies. For noninteracting nucleons with
$S_\sigma(\omega)=2\pi\delta(\omega)$ it is found to be
$\sigma_T=9C_A^2 G_{\rm F}^2T^2/\pi$. With the ansatz Eq.~(\ref{eq:3})
one finds that the cross section varies with $\Gamma_\sigma$ as shown
in Fig.~\ref{fig:reduction} where $s(x)=1$ has been chosen.

The dashed line is the tangent to the solid line at the point
$\Gamma_\sigma=0$.  It actually does not depend on the Lorentzian
ansatz for the low-$\omega$ behavior. Any modification of $S^{\rm
brems}_\sigma(\omega)$ which is chosen such that the sum rule is
obeyed yields the same result for the dashed line \cite{RSS}. It can
also be derived directly with perturbative methods \cite{Sawyer95}.

Except in the dilute-medium limit ($\Gamma_\sigma\ll T$) it is not
obvious how to compute $\Gamma_\sigma$ as a function of the medium's
temperature and density. A naive extension of a perturbative
calculation based on a one-pion exchange potential between the
nucleons indicates that in a SN core $\Gamma_\sigma\gg T$ which would
lead to a vast suppression of the neutrino opacities. Clearly, neither
the naive nor the perturbatively suppressed cross sections shown in
Fig.~\ref{fig:reduction} are adequate proxies for the true neutrino
scattering cross section in a SN core.

One would expect that the true nucleon spin-fluctuation rate is
smaller than indicated by a perturbative calculation. Sigl \cite{Sigl}
has derived an $f$-sum rule for $S_\sigma(\omega)$ which indicates
that $\Gamma_\sigma$ never exceeds a few $T$. The same conclusion is
empirically reached on the basis of the SN~1987A neutrino signal which
should have been much shorter than observed if the neutrino opacities
had been far below their ``standard'' (i.e.\ naive) values
\cite{KJR,JKRS}.

\section{Conclusion}

In a SN core, nucleon spin autocorrelations lead to a significant
reduction of the axial-vector neutrino scattering cross
section. However, the exact magnitude of this effect cannot be
calculated, at the present time, on the basis of first principles. In
addition, spin-spin correlations must be included which complicate the
problem even further. Of course, in the spirit of a dimensional
analysis the overall magnitude of the neutrino opacities in a SN core
are correctly estimated by the naive values which ignore all
correlation effects. However, the true opacities appear to be reduced
by a factor of order unity which is not calculable at the present
time. The duration of the SN~1987A signal seems to indicate that the
opacity suppression cannot have been too severe, but this conclusion
rests on the very small number of late events in the SN
signal. Significant theoretical efforts are needed to perform a
meaningful calculation of the neutrino opacities in a nuclear
medium. Of course, a statistically more significant neutrino signal
from a future galactic SN would go a long way at resolving these
problems empirically!

\section*{Acknowledgments}

Partially support by the European Union contract CHRX-CT93-0120 and by
the Deutsche Forschungsgemeinschaft grant SFB 375 is acknowledged.


\end{document}